\documentclass[12pt]{article}
\usepackage{amsfonts}
\usepackage{amssymb}
\usepackage{epsfig}
\catcode`\@=11
\newcount\hour
\newcount\minute
\newtoks\amorpm \hour=\time\divide\hour by 60\minute
=\time{\multiply\hour by 60 \global\advance\minute by-\hour}
\edef\standardtime{{\ifnum\hour<12 \global\amorpm={am}%
        \else\global\amorpm={pm}\advance\hour by-12 \fi
        \ifnum\hour=0 \hour=12 \fi
        \number\hour:\ifnum\minute<10
        0\fi\number\minute\the\amorpm}}
\edef\militarytime{\number\hour:\ifnum\minute<10
0\fi\number\minute}
\def\draftlabel#1{{\@bsphack\if@filesw {\let\thepage\relax
   \xdef\@gtempa{\write\@auxout{\string
      \newlabel{#1}{{\@currentlabel}{\thepage}}}}}\@gtempa
   \if@nobreak \ifvmode\nobreak\fi\fi\fi\@esphack}
        \gdef\@eqnlabel{#1}}
\def\@eqnlabel{}
\def\@vacuum{}
\def\marginnote#1{}
\def\draftmarginnote#1{\marginpar{\raggedright\scriptsize\tt#1}}
\def\draft{
        \pagestyle{plain}
        \overfullrule=2pt
        \oddsidemargin -.1truein
        \def\@oddhead{\sl \phantom{\today\quad\militarytime} \hfil
        \smash{\Large\sl DRAFT} \hfil \today\quad\militarytime}
        \let\@evenhead\@oddhead
        \let\label=\draftlabel
        \let\marginnote=\draftmarginnote
        \def\ps@empty{\let\@mkboth\@gobbletwo
        \def\@oddfoot{\hfil \smash{\Large\sl DRAFT} \hfil}
        \let\@evenfoot\@oddhead}
        \def\@eqnnum{(\theequation)\rlap{\kern\marginparsep\tt\@eqnlabel}%
        \global\let\@eqnlabel\@vacuum}  }

\renewcommand{\theequation}{\thesection.\arabic{equation}}
\renewcommand{\thefootnote}{\fnsymbol{footnote}}

\def\appendix#1{\addtocounter{section}{1}\setcounter{equation}{0}
\renewcommand{\thesection}{\Alph{section}}
\section*{Appendix \thesection\protect\indent \parbox[t]{11.15cm}{#1}}
\addcontentsline{toc}{section}{Appendix \thesection\ \ \ #1}}

\def \bi{\bibitem}

\def\st {\bowtie}
\def\be{\begin{equation}}
\def\ee{\end{equation}}

\def \st {\ltimes}

\catcode`\@=11

\def\bea{\begin{eqnarray}}
\def\eea{\end{eqnarray}}
\def\beann{\begin{eqnarray*}}
\def\eeann{\end{eqnarray*}}
\def\beq{\begin{equation}}
\def\eeq{\end{equation}}
\def\ba{\begin{array}}
\def\ea{\end{array}}
\def\ben{\begin{enumerate}}
\def\een{\end{enumerate}}

 \def\be{\begin{equation}}
\def\ee{\end{equation}}

\font\mybb=msbm10 at 11pt

\def\bb#1{\hbox{\mybb#1}}

\def\bR {\bb{R}}

\def \ee {\epsilon}

\def \bi{\bibitem}

\def\be{\begin{equation}}
\def\ee{\end{equation}}

\def \bi {\bibitem}

\def \nn {\nonumber}

\begin{document}
\date{July 2003}
\begin{titlepage}
\begin{center}
\vspace{2.0cm} {\Large \bf
 The holonomy of IIB  supercovariant connection}\\[.2cm]

\vspace{1.5cm}
 {\large  G. Papadopoulos
 and D. Tsimpis}

 \vspace{0.5cm}

 Department of Mathematics\\
 King's College London\\
 Strand\\
 London WC2R 2LS
\end{center}

\vskip 1.5 cm
\begin{abstract}
\noindent We show that the holonomy
of the supercovariant connection of IIB supergravity is contained in $SL(32, \bR)$.
We also find that the
 holonomy  reduces to a subgroup of
$SL(32-N)\st (\oplus^N \bR^{32-N})$  for IIB supergravity backgrounds
with $N$ Killing spinors.
We  give the necessary and sufficient conditions
 for a IIB background to admit $N$ Killing spinors.
A IIB supersymmetric  probe configuration can involve
up to 31 linearly independent planar branes and
preserves one supersymmetry.

\end{abstract}
\end{titlepage}
\newpage
\setcounter{page}{1}
\renewcommand{\thefootnote}{\arabic{footnote}}
\setcounter{footnote}{0}

\setcounter{section}{0}
\setcounter{subsection}{0}

\noindent Supersymmetric backgrounds in various supergravity
theories can be characterized by the holonomy of their
supercovariant connection. For example, the holonomy of the
supercovariant connection of maximally supersymmetric solutions
is such that the  curvature of the supercovariant connection
vanishes. The latter property has been used in \cite{gpjfof}
to classify all maximally supersymmetric solutions in eleven and ten
dimensions.
The holonomy of the supercovariant connection of a generic
eleven-dimensional supergravity background is a subgroup
of $SL(32, \bR)$ \cite{hull, tsimpis}. For backgrounds with
$N$ Killing spinors it reduces to $SL(32-N, \bR)\st (\oplus^N \bR^{32-N})$.
This has been used in \cite{tsimpis} to give the necessary and sufficient
conditions for an eleven-dimensional background to admit
$N$ Killing spinors. The holonomy of the supercovariant connection is known
for many eleven-dimensional backgrounds, see \cite{hull, duff} and references 
therein. 
For another
approach to investigating supersymmetric solutions in eleven-dimensions
see \cite{pakis}.

In this note,
we shall show that the (restricted) holonomy, ${\rm hol}({\cal D})$,
 of the supercovariant connection,
 ${\cal D}$, of IIB
supergravity is $SL(32, \bR)$, i.e. it is the same as that of the
supercovariant connection of eleven-dimensional supergravity. Consequently,
the holonomy of the supercovariant connection of a background with $N$ Killing spinors
is a subgroup of $SL(32-N, \bR)\st (\oplus^N \bR^{32-N})$. 
A IIB background admits
exactly $N$ Killing spinors iff
\be
SL(31-N, \bR)\st (\oplus^{N+1} \bR^{31-N})\nsupseteq {\rm hol}({\cal D})\subseteq SL(32-N, \bR)\st (\oplus^N \bR^{32-N})~.
\ee
Adapting the results of \cite{tsimpis} to ten dimensions, we conclude that
there is no topological obstruction for the existence of $N=22$ Killing spinors.
The first obstruction can occur for $N=23$ provided that the top cohomology class
of spacetime does not vanish.
If spacetime has topology $\bR\times \Sigma$, an obstruction can occur for $N=24$.
The holonomy of the supercovariant connections of standard and massive
IIA supergravities takes values in $SL(32, \bR)$ as well.

The Clifford algebra
 ${\rm Cliff}(\bR^{9,1})$ as a vector space is
isomorphic to $\Lambda^*(\bR^{9,1})$, ${\rm Cliff}(\bR^{9,1})=\Lambda^*(\bR^{9,1})$,
and so it has dimension $2^{10}$. (For a summary see \cite{harvey}).
The Clifford algebra as an algebra is  isomorphic to
\be
{\rm Cliff}(\bR^{9,1})=M_{32}(\bR)~,
\label{cliff}
\ee
where $M_n(\bR)$ is the space of $n\times n$ matrices with real entries.
We have
$$
{\rm Cliff}(\bR^{9,1})={\rm Cliff}^{{\rm even}}(\bR^{9,1})
\oplus {\rm Cliff}^{{\rm odd}}(\bR^{9,1})~,
$$
corresponding to the decomposition of $\Lambda^*(\bR^{9,1})$ in terms of
even- and odd-degree forms.
It is known that
\be
{\rm Cliff}^{{\rm even}}(\bR^{9,1})=M_{16}(\bR)\oplus M_{16}(\bR)
\label{ecliff}
\ee
and so $Spin(9,1)\subset {\rm Cliff}^{{\rm even}}(\bR^{9,1})=M_{16}(\bR)\oplus M_{16}(\bR)$.
${\rm Cliff}(\bR^{9,1})$
has one irreducible (pinor) representation
 of dimension 32, given by the standard action of $M_{32}(\bR)$ on
$\bR^{32}$.
The even part of the Clifford algebra has two  irreducible representations
given by the standard action of $M_{16}(\bR)$ on $\bR^{16}$, one for each
factor in  (\ref{ecliff}). These are the well-known Majorana-Weyl
spinor representations $\Delta^\pm$
of $Spin(9,1)$.
Therefore,
\be
Spin(9,1)\subset SL(16, \bR)\subset GL(16, \bR)~.
\label{subgr}
\ee
The product of two spinor representations can be decomposed
as
$$
\Delta^{\pm}\otimes \Delta^\pm= \Lambda^{1} (\bR^{9,1})\oplus \Lambda^3(\bR^{9,1})\oplus
\Lambda^{5\pm}(\bR^{9,1})~,
$$
and
$$
\Delta^{+}\otimes \Delta^-= \Lambda^{0} (\bR^{9,1})\oplus \Lambda^2(\bR^{9,1})\oplus
\Lambda^{4}(\bR^{9,1})~.
$$
In particular, we have
$$
\Delta^{\pm}\otimes \Delta^\pm=M_{16}(\bR)
$$
for all choices of representations. In what follows, we denote by $S^\pm$ the spinor bundles
associated with the representation $\Delta^\pm$, respectively.

IIB supergravity \cite{schwarz, howe} has two Killing spinor equations. One is associated
with the supersymmetry variation of the gravitino and the other
with the supersymmetry variation of the axionino-dilatino. The former
is differential and it is the parallel transport equation of the
supercovariant connection. The latter is algebraic. We shall be  concerned
mostly with the gravitino Killing spinor equation.
There are many ways to write the Killing spinor equations of IIB
supergravity. For our purpose, it is convenient to use the form given in
\cite{tomas} after adjusting for conventions\footnote{We use a mostly plus metric, $\epsilon_{01\dots 9}=1$,
$\Gamma_{11}=\Gamma^{0123456789}$ and the gamma matrices are real.}.
The gravitino Killing spinor equation in the string frame is
\be
{\cal D}_M\epsilon=0~,
\ee
where $\epsilon^{\rm t}=(\epsilon^1, \epsilon^2)$
is a chiral spinor doublet, i.e. a section
of the bundle $S^+\oplus S^+$,  ${\cal D}$ is the supercovariant connection,
\be
{\cal D}_M=\nabla_M +{1\over8} H_{MAB} \Gamma^{AB}\otimes \sigma_3+ {1\over16}
e^\phi \sum_{n=1}^5 {(-1)^{n-1}\over (2n-1)!} G_{A_1\dots A_{2n-1}}
\Gamma^{A_1\dots A_{2n-1}} \Gamma_M\otimes \lambda_n~,
\ee
$\nabla$ is the Levi-Civita connection and $\phi$ is the dilaton.
We have used the notation $\lambda_n=\sigma_1$, for $n$ even, 
and $\lambda_n=i\sigma_2$, for $n$ odd; $\{\sigma_i :
i=1,2,3\}$ are the Pauli matrices.
In addition, we have\footnote{Our conventions for a k-degree 
form are $\omega={1\over k!} \omega_{i_1\dots i_k} dx^{i_1}
\wedge\dots\wedge dx^{i_k}$.}
\bea
H&=&dB
\cr
G^{2n+1}&=&dC^{2n}- H\wedge C^{2n-2}~,
\eea
where $H$ ($B$) is the NS-NS three-(two-)form field strength 
(gauge potential) and $G^{2n+1}$ ($C^{2n}$)
are the R-R $2n+1$- ($2n$-)form field strengths (gauge potentials).
The independent RR field strengths 
are $G^1$, $G^3$ and $G^5$, where $G^9={}^* G^1$, $G^7=-{}^* G^3$,
and $G^5={}^*G_5$ is self-dual.

The supercovariant connection of IIB supergravity is a connection 
on the bundle $S^+\oplus S^+$.
This bundle can be viewed as an associated vector bundle 
$P(G)\times_\rho (\Delta^+\otimes \bR^2)$
of a principal bundle $P(G)$ with fibre group $G=H\times SL(2, \bR)$
for $H=Spin(9,1),  SL(16, \bR)$ or $GL(16, \bR)$. The representation $\rho$
 is the standard representation of
$GL(16, \bR)\times SL(2, \bR)$ on $\Delta^+\otimes \bR^2=\bR^{32}$ 
restricted to the subgroup $G$.
The bundle $S^+\oplus S^+$ is also an associated bundle of
 $SL(32, \bR)$ and $GL(32, \bR)$.
This can been  seen by observing that $Spin(9,1)$ and
 $SL(16, \bR)$ can be diagonally embedded
in $SL(32, \bR)$ and in $GL(32, \bR)$. In order to investigate
 the properties of the supercovariant
connection, 
it is best to view 
$S^+\oplus S^+$ as an associated bundle of $SL(32, \bR)$.

The holonomy ${\rm hol}({\cal D})$ of the superconnection 
${\cal D}$ can be any subgroup of $GL(32, \bR)$.
We shall find that it is a subgroup of $SL(32, \bR)$.
To show this, we have to compute the curvature
\be
{\cal R}_{MN}=[{\cal D}_M, {\cal D}_N]~
\ee
of the supercovariant derivative.
The supercovariant curvature ${\cal R}_{MN}$ can be expanded in a basis
involving gamma matrices as follows
\bea
{\cal R}_{MN}&=&\sum_{\mu=1}^3\phi_{MN}{}^{ \mu}~ 1_{16}\otimes \tau_\mu
\cr
&+& \sum _{\mu=0}^3\sum_{n=1}^2\phi_{MN}{}^{A_1\dots A_{2n} \mu}
\Gamma_{A_1\dots A_{2n}}\otimes \tau_\mu~,
\eea
where $\tau_0=1_2$, $\tau_1=\sigma_1$, $\tau_2=i\sigma_2$ and $\tau_3=\sigma_3$.
Observe that ${\cal R}$ is expanded in even powers of gamma matrices.
The coefficients $\phi$ are functions of the bosonic fields of
IIB supergravity.
As we have seen the basis involving even powers of gamma matrices
spans $M_{16}(\bR)=gl(16, \bR)$. These are tensored with the space of
$2\times 2$ matrices. Thus the curvature of the supercovariant
derivative takes values in\footnote{In general $M_n(\bR)\otimes M_m(\bR)=M_{nm}(\bR)$.}
 $$
 M_{16}(\bR)\otimes M_2(\bR)= M_{32}(\bR)~.
 $$
However the component of ${\cal R}$ along the 
generator $1_{16}\otimes 1_2$ vanishes. 
To see this observe that the only terms
that can contribute to the component of ${\cal R}$
along the generator $1_{16}\otimes 1_2$, are proportional to
\be
H_{MA_1A_2} H_{NB_1B_2} [\Gamma^{A_1A_2}, \Gamma^{B_1B_2}]\otimes 1_2
\label{fterm}
\ee
and
\be
G_{A_1\dots A_{2n-1}} G_{B_1\dots B_{2n-1}}
 [\Gamma^{A_1\dots A_{2n-1}}\Gamma_M, \Gamma^{B_1\dots B_{2n-1}}
\Gamma_N]\otimes 1_2~.
\label{sterm}
\ee
The remaining terms are always of the type 
$\Gamma^{(2n)}\otimes \tau_\mu$ for $n=0,2,4$ and $\mu=1,2,3$, where
$\Gamma^{(k)}$ denotes the skew-symmetric product of $k$ gamma matrices.
It is easy to see that (\ref{fterm}) vanishes
along $1_{16}\otimes 1_2$. The only terms in the decomposition
of (\ref{sterm}) which are along $1_{16}\otimes 1_2$ are proportional to either
\be
G_{M A_1\dots A_{2n-2}} G_{N}{}^{A_1\dots A_{2n-2}}
\ee
or
\be
g_{MN} G_{A_1\dots A_{2n-1}} G^{A_1\dots A_{2n-1}}~.
\ee
In both cases they vanish because of the anti-symmetry in the $M,N$ indices.
Thus the non-vanishing components of ${\cal R}$ are along the generators
$1_{16}\otimes \tau_i$, 
 $\Gamma^{(2)}\otimes \tau_\mu$ and $\Gamma^{(4)}\otimes \tau_\mu$;
$i=1,2,3$, $\mu=0,1,2,3$.
These
span the subspace $M^0_{32}(\bR)$ of $M_{32}(\bR)$ 
of matrices that have zero trace, and $M^0_{32}(\bR)=sl(32,\bR)$.
Thus we conclude
that
\be
{\rm hol}({\cal D})\subseteq SL(32, \bR)~.
\ee

The existence of parallel (Killing) spinors with respect to the
supercovariant derivative ${\cal D}$, i.e. spinors $\epsilon$ such that
$$
{\cal D}_M\epsilon=0~,
$$
implies
$$
{\cal R}_{MN}\epsilon=0
$$
and the reduction of the holonomy group to a subgroup of $SL(32, \bR)$.
If the background admits $N$ Killing spinors, then
$$
{\rm hol}({\cal D})\subseteq SL(32-N, \bR)\st (\oplus^N \bR^{32-N})~.
$$
The proof is similar to the one given in 
the case of eleven-dimensional supergravity.
Similarly,
a background admits precisely $N$ Killing spinors iff
\be
SL(31-N, \bR)\st (\oplus^{N+1} \bR^{31-N})
\nsupseteq {\rm hol}({\cal D})\subseteq SL(32-N, \bR)\st (\oplus^N \bR^{32-N})
~.
\label{cone}
\ee
These conditions together with the field equations of IIB supergravity 
are the {\sl necessary and sufficient}
conditions for a IIB supergravity 
background to admit $N$ Killing spinors.

Conditions (\ref{cone}) can be expressed in
 terms of conditions on the supercovariant curvature. For this
we adopt a moving frame along the parallel spinors and
introduce the basis $\{m_{ab}; a,b=1, \dots, 32\}$ 
of $sl(32, \bR)$ given for example in \cite{tsimpis}.
Since $\{1_{16}\otimes \tau_i, \Gamma^{(2)}\otimes \tau_\mu,
 \Gamma^{(4)}\otimes \tau_\mu; i=1,2,3,~ \mu=0,1,2,3\}$
is another basis of $sl(32, \bR)$, there is an 
invertible transformation  $Y$ such that
\be
\Gamma_{A_1\dots A_{2n}}\otimes \tau_\mu= Y_{A_1\dots A_{2n}, \mu}{}^{ab} m_{ab}~,
\ee
where $n$ and $\mu$ have the appropriate ranges.
Therefore we can expand
\be
{\cal R}_{MN}= {\cal R}_{MN}{}^{ab} m_{ab}~.
\ee
 The conditions for the existence
of $N$ Killing spinors are
\be
{\cal R}_{MN}{}^{a b}=0~, a=1, \dots, 32~~~~~b=1, \dots, N~.
\ee
These are similar to those found in eleven-dimensional supergravity.

{}For completeness, we give the curvature of the supercovariant derivative
of IIB supergravity in the notation of \cite{schwarz}
 to which the reader
is referred for further details. ${\cal D}_M$ is given by
\be
{\cal D}_M\epsilon:=D_M\epsilon+U_M\epsilon+V_M\epsilon^\star~,
\ee
where
\bea
D_M\epsilon&:=&
\left(\partial_M+{1\over 4}\omega_M{}^{ab}\Gamma_{ab}
-{i\over 2}Q_M\right)\epsilon\nn\\
U_M&:=&{i\kappa\over 48}\Gamma^{L_1\dots L_4}
F_{ML_1\dots L_4}\nn\\
V_M&:=&{\kappa\over 96}
\left(\Gamma_M{}^{L_1L_2L_3}
G_{L_1L_2L_3}-9\Gamma^{L_1L_2}
G_{ML_1L_2}
\right)~.
\eea
A computation yields
\be
{\cal D}_{[N}{\cal D}_{M]}\epsilon=
{1\over 2}{\cal R}_{NM}\epsilon={\cal S}\epsilon
+{\cal T}\epsilon^\star~,
\label{epicurus}
\ee
where the 16 by 16 complex matrices ${\cal S}$,${\cal T}$ are given by
\bea
{\cal S}&=&{1\over 8}R_{NM}{}^{L_1L_2}\Gamma_{L_1L_2}
-{1\over 2}P_{[N}P^\star_{M]}+
{i\kappa\over 48}\Gamma^{L_1\dots L_4}D_{[N}F_{M]L_1\dots L_4} \nn\\
&&+{\kappa^2\over 24}(
-\Gamma^{L_1L_2}F_{[N|L_1}{}^{Q_1Q_2Q_3}F_{|M]L_2 Q_1Q_2Q_3}
+{1\over 2}\Gamma^{L_1\dots L_4}F_{NML_1}{}^{Q_1Q_2}F_{L_2L_3L_4Q_1Q_2}\nn\\
&&~~~~~~~~~~~~~~~~~~~~~~~~+
{1\over 2}\Gamma_{[N}{}^{L_1L_2 L_3}F_{M]L_1}{}^{Q_1Q_2Q_3}
F_{L_2L_3Q_1Q_2Q_3}) \nn\\
&&+{\kappa^2\over 32}(
-{1\over2}G_{[N}{}^{L_1L_2}G^\star_{M]L_1L_2}
+{1\over48}\Gamma_{NM}G^{L_1L_2L_3}G^\star_{L_1L_2L_3}\nn\\
&&~~~~~~-{1\over4}\Gamma_{[N}{}^{L_1}G_{M]}{}^{L_2L_3}G^\star_{L_1L_2L_3}
+{1\over8}\Gamma_{[N|}{}^{Q}G_{Q}{}^{L_1L_2}G^\star_{|M]L_1L_2}\nn\\
&&~~~~~~+{3\over16}\Gamma^{L_1L_2}G_{NM}{}^{L_3}G^\star_{L_1L_2L_3}
-\Gamma^{L_1L_2}G_{[N|L_1}{}^{Q}G^\star_{|M]L_2Q}\nn\\
&&~~~~~~-{3\over16}\Gamma^{L_1L_2}G_{L_1L_2}{}^{Q}G^\star_{NMQ}
+{1\over16}\Gamma_{NM}{}^{L_1L_2}G_{L_1}{}^{Q_1Q_2}G^\star_{L_2Q_1Q_2}
\nn\\
&&~~~~~~-{1\over16}\Gamma^{L_1\dots L_4}G_{L_1L_2L_3}G^\star_{NML_4}
+{1\over8}\Gamma_{[N|}{}^{L_1L_2L_3}G_{L_1L_2}{}^{Q}G^\star_{|M]L_3Q}\nn\\
&&~~~~~~+{1\over4}\Gamma^{L_1\dots L_4}G_{[N|L_1L_2}G^\star_{|M]L_3L_4}
+{1\over16}\Gamma^{L_1\dots L_4}G_{NML_1}G^\star_{L_2L_3L_4}\nn\\
&&~~~~~~+{1\over4}\Gamma_{[N|}{}^{L_1L_2L_3}G_{|M]L_1}{}^{Q}G^\star_{L_2L_3Q}
+{1\over24}\Gamma_{[N|}{}^{L_1\dots L_5}G_{|M]L_1L_2}G^\star_{L_3L_4L_5}\nn\\
&&~~~~~~-{1\over48}\Gamma_{[N|}{}^{L_1\dots L_5}
G_{L_1L_2L_3}G^\star_{|M]L_4L_5}
-{1\over32}\Gamma_{NM}{}^{L_1\dots L_4}G_{L_1L_2}{}^{Q}G^\star_{L_3L_4Q}\nn\\
&&~~~~~~-{1\over288}\Gamma_{NM}{}^{L_1\dots L_6}
G_{L_1L_2L_3}G^\star_{L_4L_5L_6}
)
\label{aurelius}
\eea
and
\bea
{\cal T}&=&-{\kappa\over96}(
\Gamma_{[N}{}^{L_1L_2L_3}D_{M]}G_{L_1L_2L_3}
+9\Gamma^{L_1L_2}D_{[N}G_{M]L_1L_2})\nn\\
&&+{i\kappa^2\over 32}(
{1\over3}F_{NM}{}^{L_1L_2L_3}G_{L_1L_2L_3}
+\Gamma^{L_1L_2}F_{[N|L_1L_2}{}^{Q_1Q_2}G_{|M]Q_1Q_2} \nn\\
&&~~~~~~+{1\over3}\Gamma_{[N}{}^{Q}F_{M]Q}{}^{L_1L_2L_3}G_{L_1L_2L_3}
-{1\over2}\Gamma^{L_1\dots L_4}F_{NML_1L_2}{}^{Q}G_{L_3L_4Q}\nn\\
&&~~~~~~+{1\over2}\Gamma_{[N}{}^{L_1L_2L_3}F_{M]L_1L_2}{}^{Q_1Q_2}G_{L_3Q_1Q_2}
+{1\over4}\Gamma^{L_1\dots L_4}F_{L_1\dots L_4}{}^{Q}G_{NMQ}\nn\\
&&~~~~~~-{1\over2}\Gamma_{[N|}{}^{L_1L_2L_3}F_{L_1L_2L_3}{}^{Q_1Q_2}
G_{|M]Q_1Q_2})
~.
\label{plotinus}
\eea
Using the doubling trick
\be
i\rightarrow \pmatrix{0&1\cr -1&0},~~~~~
\epsilon=\epsilon_1+i\epsilon_2\rightarrow \pmatrix{\epsilon_1\cr 
\epsilon_2}~,
\ee
we can write (\ref{epicurus}) in terms of 32 by 32 real matrices,
\be
{1\over 2}{\cal R}_{NM}=Re({\cal S})\otimes \tau_0+Im({\cal T})\otimes \tau_1
-Im({\cal S})\otimes \tau_2+Re({\cal T})\otimes \tau_3
~.
\label{zeno}
\ee
It can be seen again   that ${\cal R}_{NM}$ takes values in $sl(32,\bR)$.

As in the case of M-theory \cite{tsimpis}, the maximal number
of {\sl linearly independent} IIB branes that can
take part in a supersymmetric probe configuration in $\bR^{9,1}$ is 31. 
This number is determined by the dimension of the Cartan subalgebra
of $sl(32, \bR)$ spanned by symmetric traceless matrices.
Such brane configurations
can be constructed by reducing a supersymmetric M-brane
configuration of M-theory with 31 linearly independent branes to IIA and 
then T-dualizing to IIB.
The subgroup $SO(32)$
of $SL(32, \bR)$
 rotates one IIB supersymmetric planar brane configuration to another. 
 Generic rotations
 by $SL(32, \bR)$ do not preserve the hermiticity of projectors.

The holonomy of the supercovariant connections of standard and massive IIA 
supergravities also takes values in $SL(32, \bR)$.
In the former case this follows either  from the similar 
result in eleven-dimensions or by directly
computing the supercovariant curvature. In the latter case it follows
from the computation of the supercovariant curvature of massive IIA 
supergravity \cite{romans}. Therefore all our results
regarding the reduction of the holonomy for backgrounds 
with N Killing spinors and the above
discussion regarding planar brane configurations apply 
to this case as well.
It is rather striking that the holonomy of the supercovariant 
connections of D=11 and type II D=10
 supergravities is $SL(32, \bR)$. However, the spaces on which 
$SL(32, \bR)$
 acts in each case are different.
   One explanation for this may be the underlying duality
 relations between 
these theories. This common property of ten- and eleven-dimensional
  supergravity theories merits
 further investigation.

\vskip 0.5cm
\noindent {\bf Acknowledgements}

\noindent 
DT thanks the European Union, grant number HPRN-2000-00122, which includes
the Department of Physics, Queen Mary College University of London,
as a sub-contractor.

\vskip 0.5cm


\end{document}